\documentclass[
    a4paper,notitlepage,
    onecolumn,
    nofootinbib,
    amsmath,amssymb
]{revtex4-2}

\usepackage{graphicx}
\graphicspath{{figures/}}

\usepackage[usenames,dvipsnames]{xcolor} 

\usepackage[
    colorlinks = true,
    urlcolor = MidnightBlue,
    linkcolor = Maroon,
    citecolor = Maroon
]{hyperref}

\hypersetup{
    pdfauthor = {A. V. Glushkov, L. T. Ksenofontov, K. G. Lebedev, A. Saburov},
    pdftitle = {Explanation to the article "On the calibration of ultra-high energy EASs at the Yakutsk array and Telescope Array"},
    pdfsubject = {ultra-high energy cosmic rays, extensive air showers},
    pdfkeywords = {Telescope Array, SD response, primary energy, spectrum}
}

\newcommand{\ESD}{E_{\text{SD}}}
\newcommand{\ESDOne}{E^1_{\text{SD}}}

\newcommand{\EFD}{E_{\text{FD}}}

\newcommand{\VEM}{\text{VEM}}
\newcommand{\MEAN}{\text{mean}}

\newcommand{\degr}{^{\circ}}

\newcommand{\eqn}[1]{(\ref{#1})}
\newcommand{\fign}[1]{Fig.~\ref{#1}}

\newcommand{\arXiv}[2] {%
    arXiv:\href{https://arXiv.org/abs/#1}{#1} [#2]%
}

\begin{document}

\title{Explanation to the article ``On the calibration of ultra-high energy EASs at the Yakutsk array and Telescope Array''}

\author{\bf A.\,V.\,Glushkov\textsuperscript{1)}}
\email{glushkov@ikfia.ysn.ru}

\author{L.\,T.\,Ksenofontov\textsuperscript{1)}}

\author{K.\,G.\,Lebedev\textsuperscript{1)}}

\author{A.\,V.\,Saburov}

\affiliation{\normalfont{\textsuperscript{1)}Yu.\,G.\,Shafer Institute of Cosmophysical Reserach and Aeronomy of Siberian branch of the Russian Academy of Sciences, 31 Lenin ave., Yakutsk, 677027, Russia}}

\maketitle

In~\cite{b:1} a critical comment is given to our preprint~\cite{b:2} where we were questioning some conclusions of the thesis~\cite{b:3}. The response~\cite{b:1} barely touches some important aspects of our paper, instead focusing on the full simulation of the surface detector (SD) performed for the Telescope Array (TA) experiment (CORSIKA code, QGSJet-II.03 model and Geant4 toolkit), and on the connection established between the calculated density of energy deposit in SD's scintillator and the energy of primary particle $\ESD$. A conclusion was made in~\cite{b:3} (incorrect in our opinion) that this energy deposit corresponds to the other value of primary energy, $\EFD$, which is estimated from EAS fluorescent light yield and is 1.27 times lesser then $\ESD$:

\begin{equation}
    \ESD = 1.27 \times \EFD\text{.}
    \label{eq:1}
\end{equation}

In~\cite{b:2} we performed similar full simulation of the TA SD and have shown that relation~\eqn{eq:1} is incorrect. According to simulation results~\cite{b:3} the chosen value of the response unit VEM (Vertical Equivalent Muon) for TA scintillator equals to 2.05~MeV. These results are presented below on figures taken from the thesis~\cite{b:3}.

\begin{figure}[!htb]
    \includegraphics[width=0.40\textwidth]{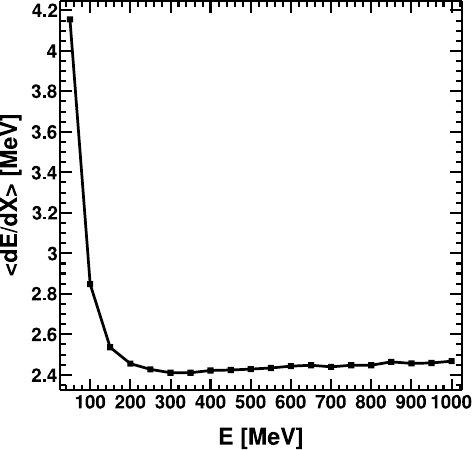}
    \caption{Mean energy deposition in a 1.2 cm thick scintillator by a vertical muon plotted versus muon kinetic energy, obtained from a Geant4 simulation. The minimum ionizing energy occurs at 300~MeV. {\sl (Figure~2.9 in \cite{b:3})}.}
    \label{f:1}
\end{figure}

\begin{figure}[!htb]
    \includegraphics[width=0.40\textwidth]{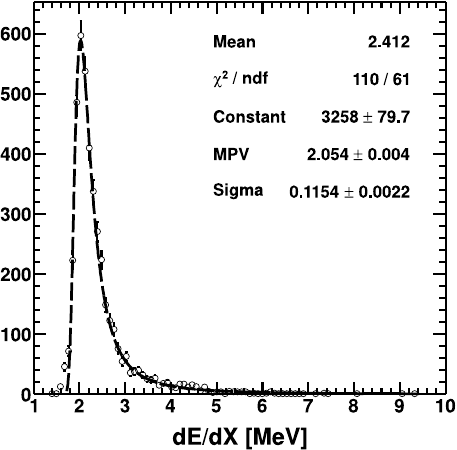}
    \caption{Energy loss histogram of a vertical 300~MeV (minimum ionizing) muon in the 1.2 cm scintillator, obtained from a Geant4 simulation. The MPV (most probable value) defines the VEM unit, which is 2.05~MeV for the TA SD. Dashed curve shows a fit to the Landau function. {\sl (Figure~2.10 in \cite{b:3})}.}
    \label{f:2}
\end{figure}

In our simulation we introduced a different unit of the energy deposited in SD during the passage of a charged relativistic particle (see equation (10) in~\cite{b:2}):

\begin{equation}
    \varepsilon = 2.05 \times 1.2 \times 1.036 \times \sec\theta~\text{MeV} = 
    2.54 \times \sec\theta~\text{MeV,}
    \label{eq:10}
\end{equation}
which, according to~\cite{b:1}, is incorrect. It is claimed that presence of the coefficients 1.2 (scintillator thickness in cm) and 1.036 (scintillator density in g/cm$^3$) in \eqn{eq:10} is wrong. A quote from~\cite{b:1}: {\sl ``It is important to note that the value 2.05 MeV is expressed in terms of the energy measured in MeV, already accounting for the thickness and density of the TA scintillator. It does not represent the energy deposit per unit length. Therefore, scaling this value by the thickness or the density of the scintillator is not meaningful''}. This statement expresses the main point of disagreement with our paper. It appears that authors of~\cite{b:1} have concluded that we used the data from \fign{f:2} where $\text{MVP} = \VEM = 2.05$~MeV. Let's discuss this in detail.

\begin{figure}
    \includegraphics[width=0.55\textwidth]{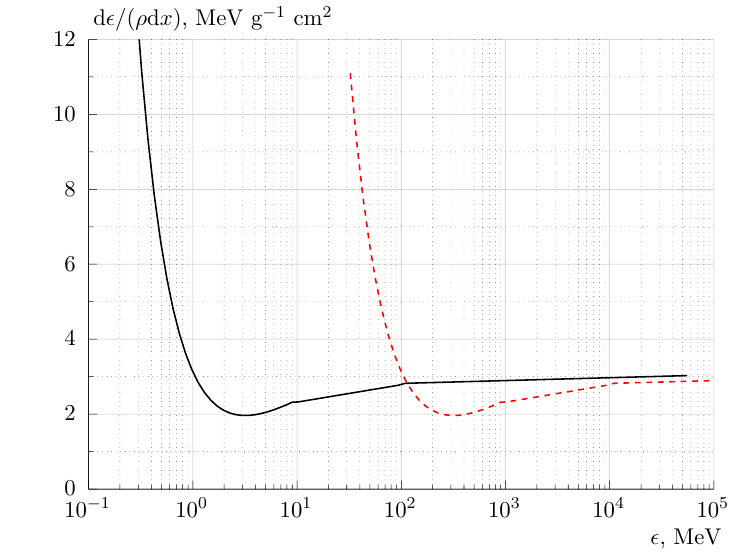}
    \caption{Differential ionization energy losses of electrons (solid curve) and muons (dashed curve) in water. {\sl (Figure~8 in \cite{b:2})}.}
    \label{f:8}
\end{figure}

For better understanding we will use Figure~8 from \cite{b:2} (see \fign{f:8} in this preprint) since all calculations of the SD response with equation \eqn{eq:10} use the data presented in this figure. First, let us note that the term VEM itself is somewhat misleading, since its name omits the contribution from electromagnetic EAS component; this is why we don't use it. On \fign{f:8} \underline{mean} differential energy losses of electrons and muons are shown, which are identical in shape. In minimum both curves have similar values of losses $\approx 2.05$~MeV (during the passage of electrons and muons through 1~cm thick layer of water). Equation \eqn{eq:10} for the TA SD was derived from it. The 2.05~MeV value in this formula is accidentally equal to the most probable value (MPV) from \fign{f:2}. It seems that this circumstance had misled the authors of~\cite{b:1}. For vertical muons with energy 300~MeV the responses $\varepsilon = 2.54$ and $\MEAN = 2.41$ (see \fign{f:1}) are close to each other. It is these values that should be considered as response units of the SD. During the passage of $N$ vertical muons with energy 300~MeV their energy losses will be different. Their summary energy losses $\sum({\rm d}E / {\rm d}x)_{N}$ measured in experiment are equal to the integrated Landau distribution, which in turn means that the mean energy deposit of one muon is

\begin{equation}
    \MEAN = \frac{1}{N}\sum\left(
        \frac{{\rm d}E}{{\rm d}x}
    \right)_{N} = 2.41~\text{MeV.}
    \label{eq:2}
\end{equation}
Hence, to obtain the number of muons that passed the detector, one should perform a reverse operation:

\begin{equation}
    N = \frac{1}{\MEAN}\sum\left(
        \frac{{\rm d}E}{{\rm d}x}
    \right)_{N}\text{.}
    \label{eq:3}
\end{equation}

The main classification parameter adopted at TA for estimation of energy $\ESD$ is the number $N$ of responses measured by SD with 3~m$^2$ area at axis distance $r = 800$~m in showers with zenith angles $\theta = 0\degr$. Calculations~\cite{b:2} have demonstrated (see section 4) that measured densities $S(800,0\degr)$ of these detectors with response unit \eqn{eq:10} are unambiguously connected with primary energy $\ESD$ via the following relation:

\begin{equation}
    \ESD = \ESDOne \times S(800,0\degr)^{1.0.25 \pm 0.010}~\text{eV,}
    \label{eq:4}
\end{equation}
where $\ESDOne = (2.29 \pm 0.08) \times 10^{17}$~eV. In the case of the value of the response unit \eqn{eq:2} the following ratio should be used in \eqn{eq:4}:

\begin{displaymath}
    \ESDOne = 2.29 \times \frac{2.54}{2.41} \times 10^{17} =
    2.17 \times 10^{17}~\text{eV.}
\end{displaymath}
From all the above we have concluded in~\cite{b:2} that results of our calculations~\cite{b:2} are virtually consistent with those presented in thesis~\cite{b:3}. \textit{This is why cosmic ray energy spectrum derived from the SD data is the \underline{correct one}}.

In TA experiment another value of response unit is officially chosen, $\VEM = 2.05$~MeV, which corresponds to the maximum of distribution presented in \fign{f:2} (MPV). It is convenient for calibration of the SD because the maximum can be easily determined by technical means. From relation \eqn{eq:3} it is evident that in this case, in order to preserve the correct number of responses $N$, one must use the relation

\begin{equation}
    N = \frac{\VEM}{\MEAN} \times N^*\text{,}
    \label{eq:5}
\end{equation}
where $N^* = \frac{\MEAN}{\VEM} \times N$ is overestimated by factor $2.41 / 2.05 \approx 1.18$. If we switch from $\VEM = 2.05$ to formula \eqn{eq:4} with $\MEAN = 2.41$, then we obtain the following relation:

\begin{equation}
    \ESD = (1.94 \pm 0.08) \times 10^{17} \times
    S(800,0\degr)^{1.025 \pm 0.010}~\text{eV.}
    \label{eq:6}
\end{equation}
It is worth noting that for some reason the TA experiment doesn't use convenient expressions similar to \eqn{eq:4}, though they are widely adopted at other world arrays.

In~\cite{b:1} no attention paid to a more important section ``{\bf 4.1. Estimation of $\ESD$}'' of~\cite{b:2}. This is the most important part of our preprint. There we claim that estimation of primary energy $\EFD$ derived from readings of optical detectors, which is unconditionally prioritized at TA, was incorrectly reduced in relation to $\ESD$ by factor $\approx 1.264$. It was reduced due to underestimation of the fraction of primary energy dispersed in the atmosphere in form of fluorescent light.


\begin{thebibliography}{3}
    \bibitem{b:1}
        J.~N.~Matthews, Y.~Tsunesada (on behalf of the Telescope Array Collaboration). Comments concerning the paper ``On the calibration of ultra-high energy EASs at the Yakutsk array and Telescope Array'' by A.V. Glushkov et al. \arXiv{2407.12892}{astro-ph.HE}.

    \bibitem{b:2}
        A.~V.~Glushkov et al. ``On the calibration of ultra-high energy EASs at the Yakutsk array and Telescope Array'', \arXiv{2404.16948}{astro-ph.HE}. To be published in Phys. At. Nucl.

    \bibitem{b:3}
        D.~Ivanov, {\sl Energy Spectrum Measured by the Telescope Array Surface Detector}, PhD thesis, New Brunswick NJ: Rutgers University (2012), DOI:~\href{https://doi.org/10.7282/T3K35SG3}{10.7282/T3K35SG3}.
\end{thebibliography}
\end{document}